\documentclass[12pt]{article}
\usepackage[dvips]{graphicx}
\usepackage{amssymb, amsmath, mathbbol}
\usepackage[centerlast,footnotesize]{caption2}

\newcommand{\lapprox}{\raisebox{-0.5ex}{$\
\stackrel{\textstyle<}{\textstyle\sim}\ $}}
\newcommand{\gapprox}{\raisebox{-0.5ex}{$\
\stackrel{\textstyle>}{\textstyle\sim}\ $}}

\usepackage{graphicx}
\newcommand{\One}{1\kern-4.5pt1}

\newcommand{\be}{\begin{equation}}
\newcommand{\ee}{\end{equation}}

\def\lesim{${\lower 2pt\hbox{$\scriptstyle
<$}\atop\raise 4pt\hbox{$\scriptstyle\sim$}}$} 
\def\grsim{${\lower2pt\hbox{$\scriptstyle >$} \atop\raise4pt\hbox 
{$\scriptstyle\sim$}}$} 

\begin{document}
\begin{center}
\begin{flushright}
June 2008
\end{flushright}
\vskip 10mm
{\LARGE
Quantum Critical Behaviour in a Graphene-like Model
}
\vskip 0.3 cm
{\bf Simon Hands$^a$ and Costas  Strouthos$^b$}
\vskip 0.3 cm
$^a${\em Department of Physics, Swansea University,\\
Singleton Park, Swansea SA2 8PP, U.K.}
\vskip 0.3 cm
$^b${\em Department of Mechanical Engineering, 
University of Cyprus,\\
Nicosia 1678, Cyprus.}
\vskip 0.3 cm
\end{center}

\noindent
{\bf Abstract:} 
We present the first results of numerical simulations of a 2+1 dimensional
fermion field theory based on a recent proposal for a model of graphene,
consisting of $N_f$ four-component Dirac fermions moving in the plane and
interacting via an instantaneous Coulomb interaction. In the strong-coupling
limit we identify a critical number of flavors $N_{fc}=4.8(2)$ separating an
insulating from a conducting phase. This transition corresponds to the location
of a quantum critical point, and we use a fit to the equation of state for the
chiral order parameter to estimate the critical exponents. Next we simulate
$N_f=2$ corresponding to real graphene, and approximately locate a transition
from strong to weak coupling behaviour.  Strong correlations are evident in the
weak-coupling regime.

                                                                                

\section{Introduction}

While there has been considerable recent interest in graphene sparked by its
discovery and subsequent experimental study~\cite{Novoselov:2005kj}, 
the remarkable properties of
electronic systems on two-dimensional honeycomb lattices have been suspected for
many years~\cite{Semenoff:1984dq}.
In brief, for a carbon monolayer having one mobile 
electron per atom,
a simple tight-binding model shows that the spectrum of low-energy excitations
exhibits a linear dispersion relation centred on zeroes located at the six 
corners of the first Brillouin
zone (eg. \cite{Gusynin:2007ix}). Using a linear transformation among 
the fields at two independent zeroes it is possible to recast the Hamiltonian
in Dirac form with $N_f=2$ flavors of 
four-component spinor $\psi$, the counting of the massless
degrees of freedom coming from 2
C atoms per unit cell $\times$ 2 zeroes per zone $\times$ 
2 physical spin components per electron. Electron propagation in the graphene
layer is thus relativistic, albeit at a speed $v_F\approx c/100$. The
implications for the high mobility of the resulting charge carriers (which may
be negatively-charged ``particles'' or positively-charged ``holes'' depending on
doping) is the source of the current excitement. The stability of the
zero-energy points is topological in origin, as emphasised in 
\cite{Creutz:2007af}.

While the above considerations apply quite generally, a realistic model of 
graphene must
incorporate interactions between charge carriers. One such model due to Son
\cite{Son:2007ja} has $N_f$ massless fermion flavors propagating in the
plane, but interacting via an
instantaneous 3$d$ Coulomb interaction. In Euclidean metric and static gauge
$\partial_0A_0=0$ the action reads
\begin{equation}
S_1=\sum_{a=1}^{N_f}\int dx_0d^2x(\bar\psi_a\gamma_0\partial_0\psi_a
+v_F\bar\psi_a\vec\gamma.\vec\nabla\psi_a+iV\bar\psi_a\gamma_0\psi_a)
+{1\over{2e^2}}\int dx_0d^3x(\partial_i V)^2,
\label{eq:model}
\end{equation}
where $e$ is the electron charge, $V\equiv A_0$ is the electrostatic potential,
 and the $4\times4$ Dirac matrices satisfy
$\{\gamma_\mu,\gamma_\nu\}=2\delta_{\mu\nu}$, $\mu=0,1,2,3$. 
In our notation
$\vec x$ is a vector in the 2$d$ plane while the index $i$ runs over all three
spatial directions. Within the graphene layer, classical
propagation of the potential 
is obtained by integrating over
the perpendicular coordinate, yielding
\begin{equation}
D_0(p)={e^2\over{2\vert\vec p\vert}}.
\end{equation}
To proceed, assume a large-$N_f$ limit so that the dominant quantum correction
$\Pi(p)$
comes from a vacuum polarisation fermion -- antifermion loop. The resummed
$V$ propagator becomes
\begin{equation}
D_1(p)=(D_0^{-1}(p)-\Pi(p))^{-1}
=\left({{2\vert\vec p\vert}\over e^2}+{N_f\over8}{{\vert\vec
p\vert^2}\over{(p^2)^{1\over2}}}\right)^{-1},
\label{eq:D_gr}
\end{equation}
where $p^2=(p_0,\vec p)^2\equiv p_0^2+v_F^2\vert\vec p\vert^2$. 
In either the strong
coupling or large-$N_f$ limits $D_1(p)$ is thus dominated by the quantum
correction, the relative importance of the original Coulomb interaction being
governed by a parameter $\lambda\equiv\vert\Pi/D_0\vert_{p_0=0}$. Restoring SI 
units, we obtain
\begin{equation}
\lambda={{e^2N_f}\over{16\varepsilon_0\hbar v_F}}\simeq1.7N_f.
\end{equation}

The form of the interaction (\ref{eq:D_gr}) means that analytic methods are
trustworthy in the large-$N_f$ regime. 
For instance, in the strong coupling limit $e^2\to\infty$\footnote{In
experiments
it is only possible to {\it reduce\/} 
the effective electron charge by mounting the graphene
layer on a dielectric substrate.}
we expect a modification of the dispersion relation, such that the fermion
energy is related to momentum via $\omega\propto p^z$, where $z$ is a dynamical
critical exponent predicted to take the value 
$z\simeq1-{4\over{\pi^2N_f}}\approx0.8$ for $N_f=2$
\cite{Son:2007ja}. Ref.~\cite{Son:2007ja} in addition
discusses the phase diagram of the graphene model
(\ref{eq:model}) in the $(N_f,e^{-2})$ plane, and raises the possibility
of symmetry breaking due to non-perturbative $N_f^{-1}$ effects.
The symmetry breaking, due to the spontaneous condensation of particle - hole
pairs, is signalled by an order parameter $\langle\bar\psi\psi\rangle\not=0$ --
in relativistic field theory this is known as ``chiral symmetry breaking''.
Physically the most important outcome is the generation of a gap in the fermion
spectrum, implying the model describes an insulator. Son postulates that this
insulating phase exists in the corner of the phase diagram corresponding to 
large $e^2$ and small $N_f$, and in particular that the insulator-conductor
phase transition taking place at $N_f=N_{fc}$ in the strong-coupling limit
$e^2\to\infty$ is a novel quantum critical point. The value of $N_{fc}$, and the
issue of whether it is greater than of less than the physical value $N_f=2$,
must be settled by a non-perturbative calculation. A recent estimate, obtained
by a renormalisation group treatment of radiatively-induced four-fermion contact
interactions, is $N_{fc}=2.03$ \cite{Drut:2007zx}. 

The proposed physics is very reminiscent of another 2+1$d$ 
fermion model, this time relativistically covariant, namely the Thirring model 
with action
\begin{equation}
S_{Th}
=\sum_{a=1}^{N_f}\int dx_0d^2x\left[\bar\psi_a\gamma_\mu\partial_\mu\psi_a
+{g^2\over2}(\bar\psi_a\gamma_\mu\psi_a)^2\right],
\label{eq:Thir}
\end{equation}
with $\mu=0,1,2$,
particularly once we insist on units such that $v_F=1$. Note that in contrast to
the graphene model the coupling $g^2$ has mass dimension -1. 
Once again, the model
is analytically tractable at large $N_f$, but exhibits spontaneous chiral 
symmetry 
breaking leading to gapped fermions
at small $N_f$ and large $g^2$ \cite{DelDebbio:1997dv,DelDebbio:1999xg}.
Arguably the Thirring model is the simplest field theory of fermions requiring a
computational solution:
the location of the phase transition at $N_f=N_{fc}$ in the strong coupling
limit has recently been determined by lattice simulation to be 
$N_{fc}=6.6(1)$~\cite{Christofi:2007ye}. The apparent similarity of the two
systems has led us to propose a Thirring-like model pertinent to graphene, with
action
\begin{equation}
S_2=\sum_{a=1}^{N_f}\int dx_0d^2x\left[\bar\psi_a\gamma_\mu\partial_\mu\psi_a
+{g^2\over2}(\bar\psi_a\gamma_0\psi_a)^2\right].
\label{eq:Thirgr}
\end{equation}
The only difference with (\ref{eq:Thir}) is that the contact interaction is now
only between the time-like components of the fermion current, so that the model
is no longer covariant.

The traditional way to proceed is to introduce an auxiliary boson field $V$.
The resulting action
\begin{equation}
S^\prime_2
=\sum_{a=1}^{N_f}\int dx_0d^2x\left[\bar\psi_a\gamma_\mu\partial_\mu\psi_a
+iV\bar\psi_a\gamma_0\psi_a+{1\over{2g^2}}V^2\right]
\label{eq:ThirV}
\end{equation}
reproduces the identical dynamics as (\ref{eq:Thirgr})
once $V$ is integrated out.
As for (\ref{eq:model}) we assume a large-$N_f$ limit to
estimate the dominant vacuum polarisation correction; the resultant propagator
for $V$ is
\begin{equation}
D_2(p)
=\left({1\over g^2}+{N_f\over8}{{\vert\vec
p\vert^2}\over{(p^2)^{1\over2}}}\right)^{-1}.
\label{eq:D_Th}
\end{equation}
In the strong-coupling or large-$N_f$ limits, 
$D_2$ coincides with $D_1$ (\ref{eq:D_gr}),
implying that the fermion interactions are equivalent. 
It is also the case that $\lim_{p\to\infty}
D_2(p)=\lim_{\lambda\to\infty}D_1(p)$.
This last limit is
important because critical behaviour in the Thirring model (\ref{eq:Thir})
is governed by a 
UV-stable fixed point of the renormalisation group~\cite{DelDebbio:1997dv}.
We aniticipate that the model (\ref{eq:Thirgr}) is similar and
expect its predictions, in
particular for critical behaviour such as the
value of $N_{fc}$, 
to be generally valid for Son's model (\ref{eq:model}) in the limit of large
$\lambda$. 

In the following section we present a version of the action (\ref{eq:ThirV}) 
discretised
on a spacetime lattice, and outline how its 
dynamics can be investigated by standard Monte Carlo
simulation techniques. To our knowledge this paper is the first to apply lattice
gauge theory techniques to graphene.
In this first paper we focus exclusively on the
equivalent of the order
parameter $\langle\bar\psi\psi\rangle$ in the $(N_f,g^{-2})$ plane.
In Sec.~\ref{sec:SC} we explore the strong coupling limit and identify 
the critical flavor number $N_{fc}$, and then attempt to characterise the
transition from insulator to conductor by studying the critical equation of
state using finite volume scaling.  In Sec.~\ref{sec:Nf=2} we switch attention
to the physical case $N_f=2$, and present results from a study of
$\langle\bar\psi\psi\rangle$ as a function of $g^2$. A brief discussion of the
implications for the graphene model of \cite{Son:2007ja} follows.

\section{Lattice Model and Simulation}

The lattice model studied in this paper is closely related to the lattice
Thirring model studied in \cite{DelDebbio:1997dv}. It is written in terms
of staggered lattice fermions, ie. 
single-component Grassmann fields $\chi,\bar\chi$ defined on the sites
$x$ of a three-dimensional cubic lattice, by the action
\begin{eqnarray}
S_{latt}={1\over2}\sum_{x\mu a}\bar\chi_{ax}\eta_{\mu x}
\bigl[(1+\delta_{\mu 0}\sqrt{{2g^2}\over N}e^{iV_x})
\chi_{a x+\hat\mu}-
(1+\delta_{\mu 0}\sqrt{{2g^2}\over N}e^{-iV_{x-\hat0}})
\chi_{a x-\hat\mu}\bigr]
\nonumber\\
+m\sum_{x a}\bar\chi_{ax}\chi_{ax}.
\label{eq:latt}
\end{eqnarray}
The flavor indices $a=1,\dots,N$. The sign factors
$\eta_{x\mu}\equiv(-1)^{x_0+\cdots+x_{\mu-1}}$ ensure that in the
long-wavelength limit the the first (antihermitian) term in $S_{latt}$ describes
the Euclidean propagation of $N_f=2N$ flavors of relativistic fermion described
by four-component spinors
\cite{BB}. The fermion mass term proportional to 
$m$ has been added
to provide a IR regulator for modes which would otherwise be massless; beyond
the usual critical slowing down, it is
important to stress that simulations directly in the limit $m\to0$ present
severe technical difficulties. The hopping terms in $S_{latt}$ 
involve the auxiliary boson field $V_x$ which is formally defined on the {\it
timelike\/} links connecting sites $x$ with $x+\hat0$. The $N$-dependence 
in the
kinetic terms is conventional, and retained 
to ensure continuity with the studies of
\cite{DelDebbio:1997dv,DelDebbio:1999xg,Christofi:2007ye}. 
In order to compare with the formulation of the
previous section the rescaling $g^2\to Ng^2$ is required.

It can be shown that the action (\ref{eq:latt}) can be reexpressed in terms of 
four-component spinors in a form similar but not identical to (\ref{eq:Thirgr}).
For full details of the relation between the two actions 
we refer the reader to \cite{DelDebbio:1997dv}. Here we merely
note that the
spontaneous generation of a condensate $\langle\bar\chi\chi\rangle\not=0$ in the
lattice model results in 
a chiral symmetry breaking pattern  U($N)\otimes$U($N)\!\to$U($N$), 
whereas in the continuum model (\ref{eq:Thirgr})
$\langle\bar\psi\psi\rangle\not=0$ breaks U($2N_f)\!\to$U($N_f)\otimes$U($N_f$)
~\cite{Gusynin:2007ix}.
A term proportional to $m$ explicitly breaks the symmetry in either case. It is
plausible that the effective global symmetry of the lattice model enlarges in
the continuum limit, and the correct continuum pattern recovered. In what
follows we will assume
that the chiral symmetry breaking described by
$\langle\bar\chi\chi\rangle\not=0$ is equivalent to the metal-insulator
transition.

The novelty of \cite{Christofi:2007ye} was the first study of the Thirring model
(\ref{eq:Thir}) by lattice means 
in the strong-coupling limit $g^2\to\infty$. Since we aim to repeat the strategy
here we discuss how this was done. First, note that the vacuum
polarisation calculation leading to the results (\ref{eq:D_gr},\ref{eq:D_Th})
does not go through in quite the same way for the lattice regularised model
(\ref{eq:latt}); rather, there is an additive correction which is
momentum independent and UV-divergent:
\begin{equation}
\Pi^{latt}(p)=\Pi^{cont}(p)+g^2J(m),
\end{equation}
where $J(m)$ comes from incomplete cancellation of a lattice tadpole diagram 
~\cite{DelDebbio:1997dv}. This extra divergence not
present in the continuum treatments can be absorbed by a wavefunction
renormalisation of $V$ and a coupling constant renormalisation
\begin{equation}
g_R^2={g^2\over{1-g^2J(m)}}.
\label{eq:gR}
\end{equation}
In the large-$N_f$ limit
we thus expect to find the strong coupling limit of the lattice model at
$g_R^2\to\infty$ implying $g^2\to g^2_{\rm lim}=J^{-1}(m)$.
For $g^2>g^2_{\rm
lim}$ $D_{latt}(p)$ becomes negative, and $S_{latt}$ no longer describes a
unitary theory.

Away from the large-$N_f$ limit, where chiral symmetry may be spontaneously
broken, there is no analytical criterion for identifying
$g^2_{\rm lim}$; however in this case a numerical calculation of
$\langle\bar\chi\chi\rangle$ shows a clear peak at $g^2=g^2_{\rm peak}$,
whose location is approximately independent of both volume and $m$, indicating
an origin at the UV scale \cite{Christofi:2007ye}.
\begin{figure}[!hbtp]
    \centering
    \includegraphics[width=13.0cm]{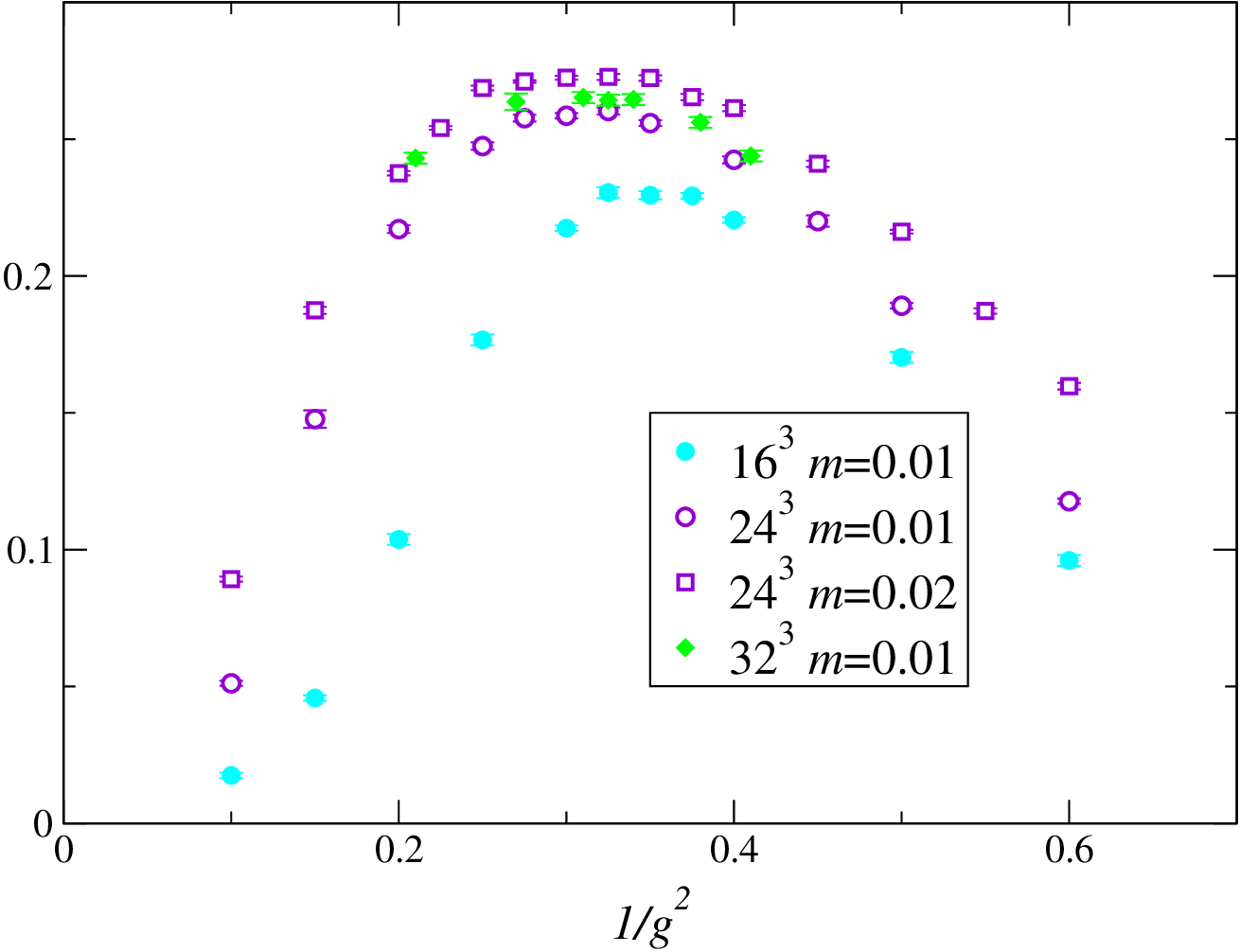}
    \caption{$\langle\bar\chi\chi\rangle$ vs. $1/g^2$ for $N_f=2$.}
   \label{gr:beta_peak_Nf2}
\end{figure}
Fig.~\ref{gr:beta_peak_Nf2} exemplifies this behaviour in the 
model (\ref{eq:latt}) with $N_f=2$ on system
volumes $L^3$: for $L\geq24$ we identify $1/g^2_{\rm peak}\simeq0.3$. Since for
orthodox chiral symmetry breaking the magnitude of the condensate is expected to
increase monotonically with the coupling strength, we interpret the peak as the
point where unitarity violation sets in, ie. 
$g^2_{\rm lim}\approx g^2_{\rm peak}$.
In the next section, we shall use simulations performed at $g^2=g^2_{\rm peak}$
to explore the strong coupling limit, and find evidence for a chiral 
symmetry restoring phase transition at a well-defined $N_{fc}$.

Writing the action (\ref{eq:latt}) in the form
$\bar\chi_iM_{ij}\chi_j$, it is possible to integrate out the fermion fields 
analytically to yield the path integral
\begin{equation}
{\cal Z}_{latt}=\int{\cal D}V(\mbox{det}M[V;m])^N.
\label{eq:FPI}
\end{equation}
Techniques to simulate the physics described by (\ref{eq:FPI}) typically
proceed by evolving a boson configuration $\{ V\}$ through a 
fictitious simulation
time using a quasi-Hamiltonian dynamics in which quantum effects are
incorporated via periodic stochastic refreshments. We implement this using the 
hybrid molecular dynamics (HMD) algorithm~\cite{HMD}.
The key step in the evolution involves the calculation of 
a force 
\begin{equation}
-{{\delta S}\over{\delta V}}=N{\rm tr}M^{-1}{{\delta M}\over{\delta V}}.
\end{equation} 
Since the force can be calculated for arbitrary $N$, it is possible to simulate
the dynamics for non-integer $N$, which is equivalent to regarding the path
integral (\ref{eq:FPI}) as the fundamental definition of the model.
Of course, only for integer $N$, and therefore for even integer $N_f$, is it
possible to express the theory as a local action in the fermion variables 
$\chi,\bar\chi$.

In the simulations described in this paper we used a HMD
algorithm to perform simulations with arbitrary $N_f$. In principle,
this method is not exact in the sense that results have a systematic dependence 
on the discrete timestep $\Delta\tau$ used to integrate the HMD equation of
motion. 
We have used $\Delta\tau=0.0025$ on the smallest systems ($16^3$, $24^3$,
$32^3$), $\Delta\tau=0.00125$ on $16^2\times48$ and $\Delta\tau=0.000625$ on 
$16^2\times64$;
in all cases 
we checked that the resulting systematic error is smaller than the statistical 
error.
The mean trajectory length $\bar\tau=1.0$, and $\langle\bar\chi\chi\rangle$
measured using 10 stochastic estimators after every trajectory. Roughly 200-400
trajectories were generated for $16^2\times48,64$, 600 for $24^3,32^3$ and
$O(1000)$ for $16^3$.
Further details of the numerical methods used can be found in 
\cite{DelDebbio:1997dv,DelDebbio:1999xg}.

\section{Results}

We performed simulations on system volumes 
$L_s^2\times L_t=16^3$, $24^3$ and $32^3$, using
fermion masses $m=0.01,\ldots,0.04$. Because
the action (\ref{eq:latt}) does not treat spacelike and timelike directions
equivalently, we also found it useful to explore the consequences of
independently varying $L_s$ and $L_t$, and thus in addition studied
$16^2\times48, 64$; $24^2\times32, 48$; and $32^2\times24$. As we shall see, the
aniostropic nature of the model's dynamics results in the systematic 
effects due to finite $L_t$ being much more important than those due to
finite $L_s$.
The only observable
discussed in this initial study is the chiral condensate 
$\langle\bar\chi\chi\rangle\equiv\langle{\rm tr}M^{-1}\rangle$.

\subsection{Strong Coupling Limit}
\label{sec:SC}

\begin{figure}[!hbtp]
    \centering
    \includegraphics[width=13.0cm]{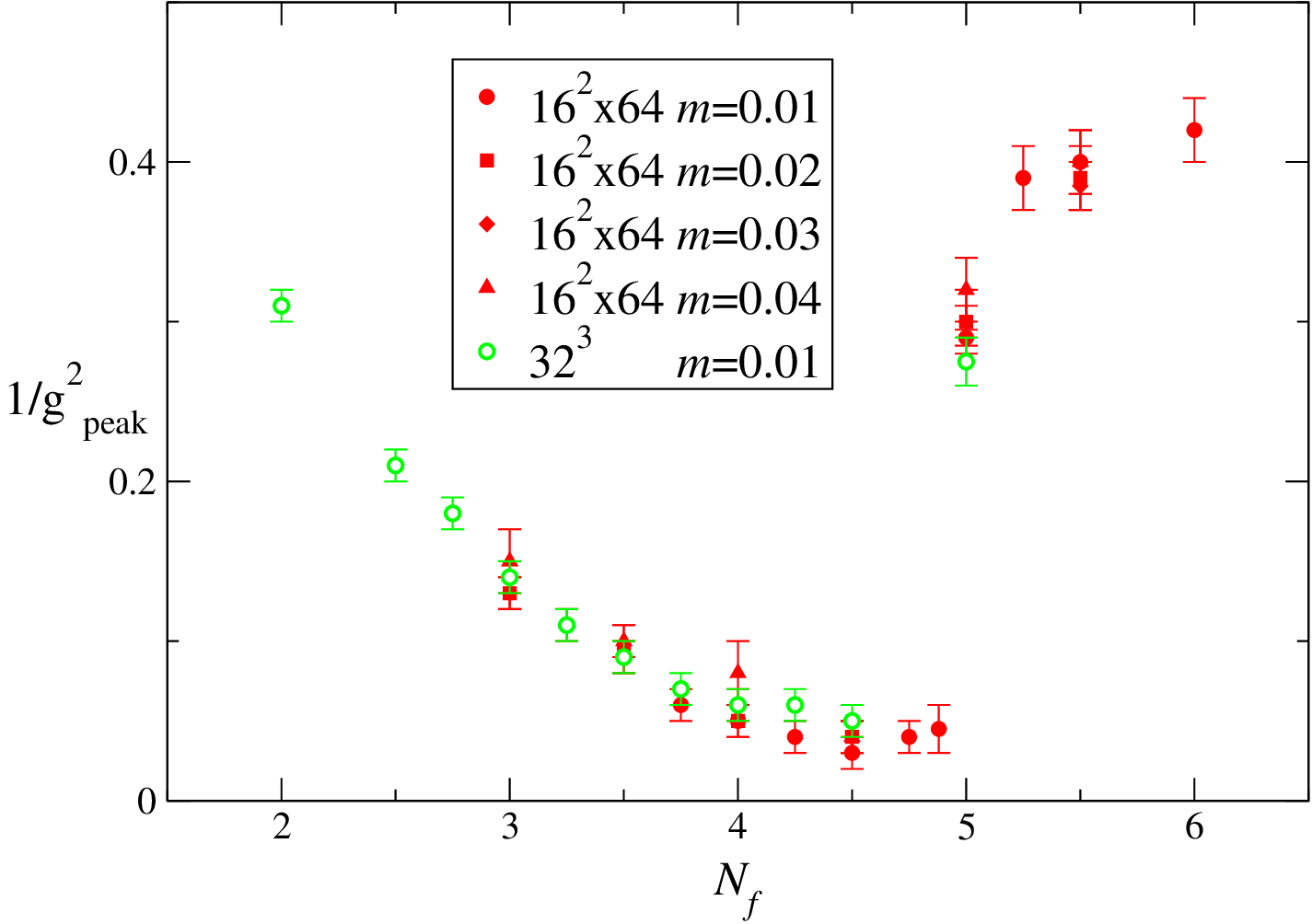}
    \caption{$1/g^2_{\rm peak}$ vs. 
$N_f$.}
   \label{gr:beta.peak}
\end{figure}
As described above, to explore the strong coupling limit $g^2_R\to\infty$ 
we made the ansatz $g^2=g^2_{\rm peak}$, where $g^2_{\rm peak}$ denotes the
location of the peak in $\langle\bar\chi\chi\rangle$ at given $N_f$.
Fig.~\ref{gr:beta.peak} shows $1/g^2_{\rm peak}(N_f)$ for some representative 
lattice volumes and fermion masses, confirming that its value, 
arising as it does from UV lattice artifacts, is to good
approximation volume and mass independent. The behaviour is qualitatively
similar to that
found for the strong-coupling Thirring model shown in Fig.~3 of
\cite{Christofi:2007ye}. We see that $1/g^2_{\rm peak}$ decreases as $N_f$
increases from 2 to 4.75, at which point the curve reaches a minimum. There is 
then a steep increase at $N_f\approx4.9$ followed by a levelling off,
implying a significant change in the 
model's strong coupling behaviour. Our interpretation, to be supported
below by a study of the equation of state, 
is that for $N_f$ below the change the model is
in a chirally broken phase, and that above the change chiral symmetry is
restored, implying $\lim_{m\to0}\langle\bar\chi\chi\rangle=0$. Using this
criterion we identify the critical flavor number for chiral symmetry restoration
in the strong coupling limit, with a conservative error, 
as 
\begin{equation}
N_{fc}=4.8(2).
\label{eq:Nfc_beta.peak}
\end{equation}
 
\begin{figure}[!hbtp]
    \centering
    \includegraphics[width=13.0cm]{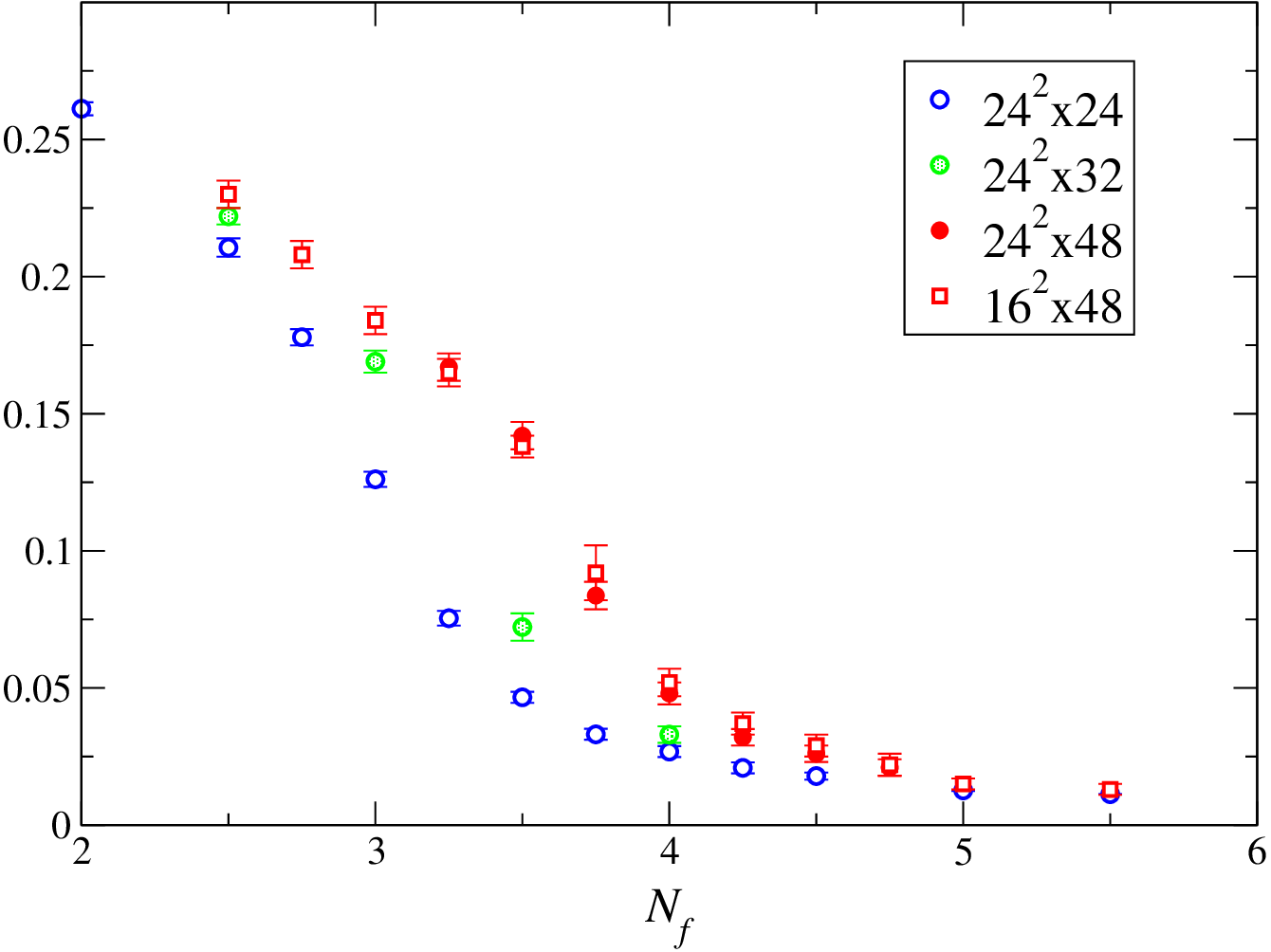}
    \caption{$\langle\bar\chi\chi\rangle$ vs. 
$N_f$ at $m=0.01$ on various lattice volumes.}
   \label{gr:volume_effects}
\end{figure}
In the rest of this section we analyse data taken
at $g^2=g^2_{\rm peak}$ in an attempt to determine the critical 
equation of state
$\langle\bar\chi\chi(m,N_f)\rangle$ in the strong coupling limit. 
Note that typically 4 -- 6 independent simulations were used to identify
$g^2_{\rm peak}$ for each $N_f$.
In the
thermodynamic zero temperature limit, 
a simple ansatz for the approximate scaling behaviour close to a quantum 
critical point is given by
\begin{equation} 
m=A(N_f-N_{fc})\langle\bar\chi\chi\rangle^p+B\langle\bar\chi\chi\rangle^\delta,
\label{eq:power}
\end{equation}
whence $\langle\bar\chi\chi\rangle\vert_{N_f=N_{fc}}\propto m^{1\over\delta}$
and the conventional exponent in
$\langle\bar\chi\chi\rangle\vert_{m=0}\propto(N_{fc}-N_f)^\beta$ for
$N_f<N_{fc}$
is given
by $\beta=(\delta-p)^{-1}$. We must, however, take account of the fact that our
data is taken on finite systems.
Fig.~\ref{gr:volume_effects} shows data for $m=0.01$ from various lattices: 
a comparison between $16^2\times48$ and $24^2\times48$
demonstrates that the
dominant finite volume effects are due to varying $L_t$, while the effects of
finite $L_s$ are negligible for $L_s\geq16$.

The theory of finite volume effects in models such as (\ref{eq:latt}) with 
anisotropic correlations is outlined in \cite{BW}. Near a critical point 
it is possible in principle to distinguish two correlation lengths $\xi_s$ and
$\xi_t$, each diverging with a distinct critical exponent $\nu_s$ , $\nu_t$ as
$N_f\to N_{fc}$. In $d$ spacetime dimensions
these are related to conventionally-defined exponents governing
scaling of the order parameter and its associated susceptibility via a
generalised hyperscaling relation
\begin{equation}
\nu_t+(d-1)\nu_s=\gamma+2\beta.
\label{eq:BW1.10}
\end{equation}
Motivated by Fig.~\ref{gr:volume_effects}, in our analysis we take a pragmatic 
	approach and assume all volume effects are due to finite $L_t$. 
The ansatz for
the modified equation of state, inspired by a renormalisation group analysis
\cite{DelDebbio:1997dv} is then
\begin{equation}
m=A[(N_f-N_{fc})+CL_t^{-{1\over\nu_t}}]\langle\bar\chi\chi\rangle^p
+B\langle\bar\chi\chi\rangle^\delta;
\label{eq:fvs}
\end{equation}
we fit this form to our dataset with a least squares fit. Our complete dataset
contains 124 data taken at various $N_f$, $m$, $L_s$, $L_t$ (recall the value 
of $1/g^2_{\rm peak}$ must be independently determined for each parameter set,
so the simulation effort involved is considerable; approximately 100 000
processor hours using 2.4GHz Opterons were required).

Experience with 
previous models shows that the fitted equation of state is very sensitive to
assumptions made about the scaling window (ie. the ranges of $N_f$ and $m$ to
include in the fit), and the smallest volume to include in the scaling ansatz
(\ref{eq:fvs}). For this reason we judge it best to present a compilation of
different fits in Table~\ref{tab:eos}.
\begin{table}[h]
\setlength{\tabcolsep}{0.4pc}
\hspace{-20mm}
\begin{tabular}{|ll|llllllll|}
\hline
fit & \# & $A$ & $B$ & $N_{fc}$ & $\delta$ & $p$ & $C$ & $\nu_t$ 
& $\chi^2$/dof \\
\hline
Power,  & 28 & 0.31(3) & 41.5(15) & 3.81(3) & 3.96(3) 
& 0.87(3) & $-$ & $-$ & 
4.8 \\
Power, $m\geq0.02$ & 18 & 0.30(7) & 87(55) & 3.87(9) & 4.4(4) 
& 0.82(6) & $-$ & $-$ &
5.9 \\
Power, $m\geq0.03$ & 12 & 2.1(10) & 3800(18) & 4.3(1) & 6.0(1) 
& 1.3(1) & $-$ & $-$ &
6.4 \\
\hline
FVS, $m=0.01$ & 53 & 1.5(7) & 63(22) & 4.60(15) & 3.9(1) & 1.3(1) & 9.7(10) &
1.7(2) & 4.0 \\
as above $L_t\geq24$ & 48 & 4.5(32) & 161(97) & 4.95(16) & 4.0(1) 
& 1.6(2) & 7.9(5) &  2.1(2) & 4.0 \\
as above $N_f<4.5$ & 46 & 712(100) & 2.3(3)$\times10^4$ & 4.46(9) & 5.25(4)
& 3.00(4) & 15(3) & 1.2(1) & 5.4  \\
FVS, all $m$ & 96 & 0.23(1) & 19(2) & 3.85(4) & 4.03(8) & 0.88(1) & 17.5(17) &
1.10(5) & 6.3 \\
as above $L_t\geq32$ & 70 & 0.20(1) & 10.5(1) & 4.7(3) & 3.6(1) & 0.82(2)
& 6.0(6) & 2.6(8) & 5.1 \\
as above $N_f\geq3$ & 60 & 0.21(1) & 237(106) & 4.6(7) & 5.5(3) & 0.86(2)
& 8.1(34) & 2.1(1.1) & 3.1 \\
$N_f\geq3$ $L_t\geq24$ & 75 & 0.21(1) & 352(137) & 4.1(1) & 5.8(2) & 0.87(2)
& 15(4) & 1.3(2) & 3.4 \\
FVS, $m\geq0.02$ & 43 & 0.19(2) & 10.4(18) & 3.76(11) & 3.55(13) & 0.78(3) & 
24(15) & 0.9(2) & 7.3 \\
as above $L_t\geq32$ & 32 & 0.16(2) & 6.5(14) & 3.9(4) & 3.21(15) & 0.72(4) & 
12(19) & 1.2(9) & 6.4 \\
\hline
\end{tabular}
\caption{Various fits to the Equations of State (\ref{eq:power}) ``Power'' and
(\ref{eq:fvs}) ``FVS''}
\smallskip
\label{tab:eos}
\end{table}
We tried fits to both the ``Power'' equation of state (\ref{eq:power}), 
with 5 free
parameters, using data
from a single lattice size $16^2\times64$, and fits using the finite-$L_t$
``FVS'' scaling form (\ref{eq:fvs}) with 7 free 
parameters.\footnote{Note it is not
possible to use hyperscaling to constrain the value of $\nu_t$, as done in
\cite{DelDebbio:1997dv,DelDebbio:1999xg,Christofi:2007ye}.} 
In the latter case data with $N_f\geq5$ was 
excluded from the fit because their small errorbars destabilised the
fits; since this data probably comes from the chirally symmetric phase there may
be a small systematic error in the identification of $g^2_{\rm peak}$ across the
transition.

Fits to (\ref{eq:power}) favour $N_{fc}\simeq3.8 - 4$, 
and $p\simeq0.9$.  These values are also favoured by the most
comprehensive FVS fit to the 96 datapoints with $N_f<5$. 
There is no evidence that discarding $m=0.01$ data, which may be most prone to 
finite volume artifacts, 
improves any of the fits. On the other hand, discarding $L_t=16$ and perhaps
$L_t=24$ does have a significant effect on the fitted values of $N_{fc}$, $p$
and $\nu_t$ in the FVS fits. 
In these cases the fitted
$\delta\approx4$. 
However, 
once data with extremal values of $N_f$ is excluded, on the assumption that
they lie outside the scaling window, the fitted values of 
$\delta$ rise to $\gapprox5$. In almost all cases the fitted value of $\nu_t$ 
exceeds 1, though often not by a statistically significant margin.

\begin{figure}[ht]
    \centering
    \includegraphics[width=13.0cm]{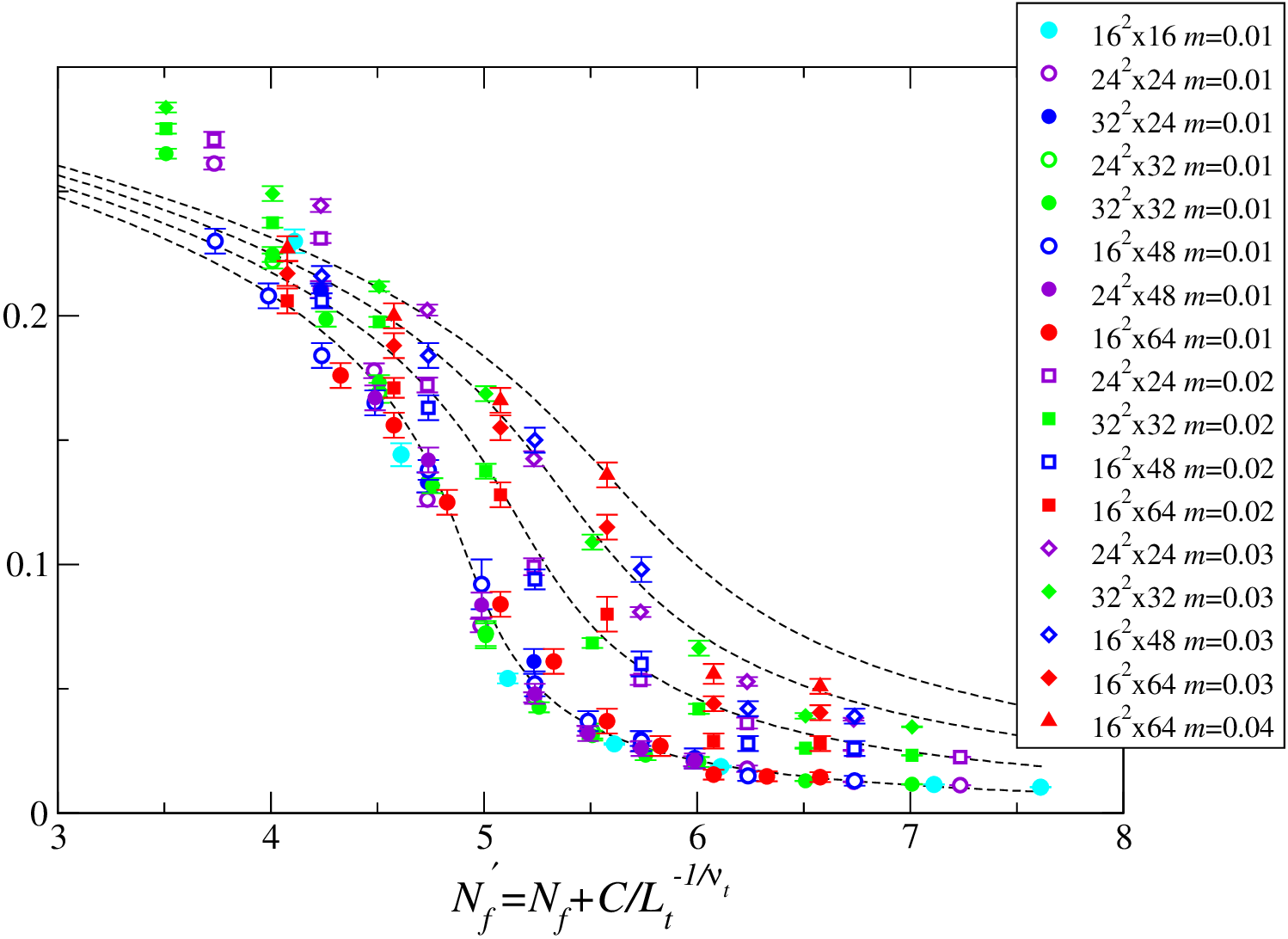}
    \caption{Finite volume scaling fit to (\ref{eq:fvs}) to data with
$m=0.01$ (circles), 0.02 (squares), 0.03 (diamonds) and 0.04 (triangles),
in terms of
$N_f^\prime$.} 
   \label{gr:fvsfit}
\end{figure}
Our favourite fit, yielding the smallest $\chi^2$/dof, emerges from the 60
datapoints with $N_f\in[3,5)$ and $L_t\geq32$. 
Another reason for preferring
this is that the fitted $N_{fc}$ is consistent with the value
(\ref{eq:Nfc_beta.peak}) coming from the behaviour of $g^2_{\rm peak}(N_f)$,
which could be regarded as an additional constraint on the global fit.
The fit is plotted in 
Fig.~\ref{gr:fvsfit} 
in terms of the control parameter in the thermodynamic limit
$N_f^\prime=N_f+CL_t^{-{1\over\nu_t}}$, so that data
with differing $L_t$ should collapse onto a single curve for each value of $m$.

To summarise: this ``best''
fit provides a reasonable description of the data in the window $4.5\lapprox
N_f^\prime\lapprox6$, in
particular for the smallest mass $m=0.01$; fits of the form
(\ref{eq:fvs}) are capable of yielding a fitted $N_{fc}$ consistent with
(\ref{eq:Nfc_beta.peak}); and the preferred value of $\nu_t\approx2$ once this
consistency criterion is applied.

\subsection{$N_f=2$}
\label{sec:Nf=2}

\begin{figure}[!hbtp]
    \centering
    \includegraphics[width=14.0cm]{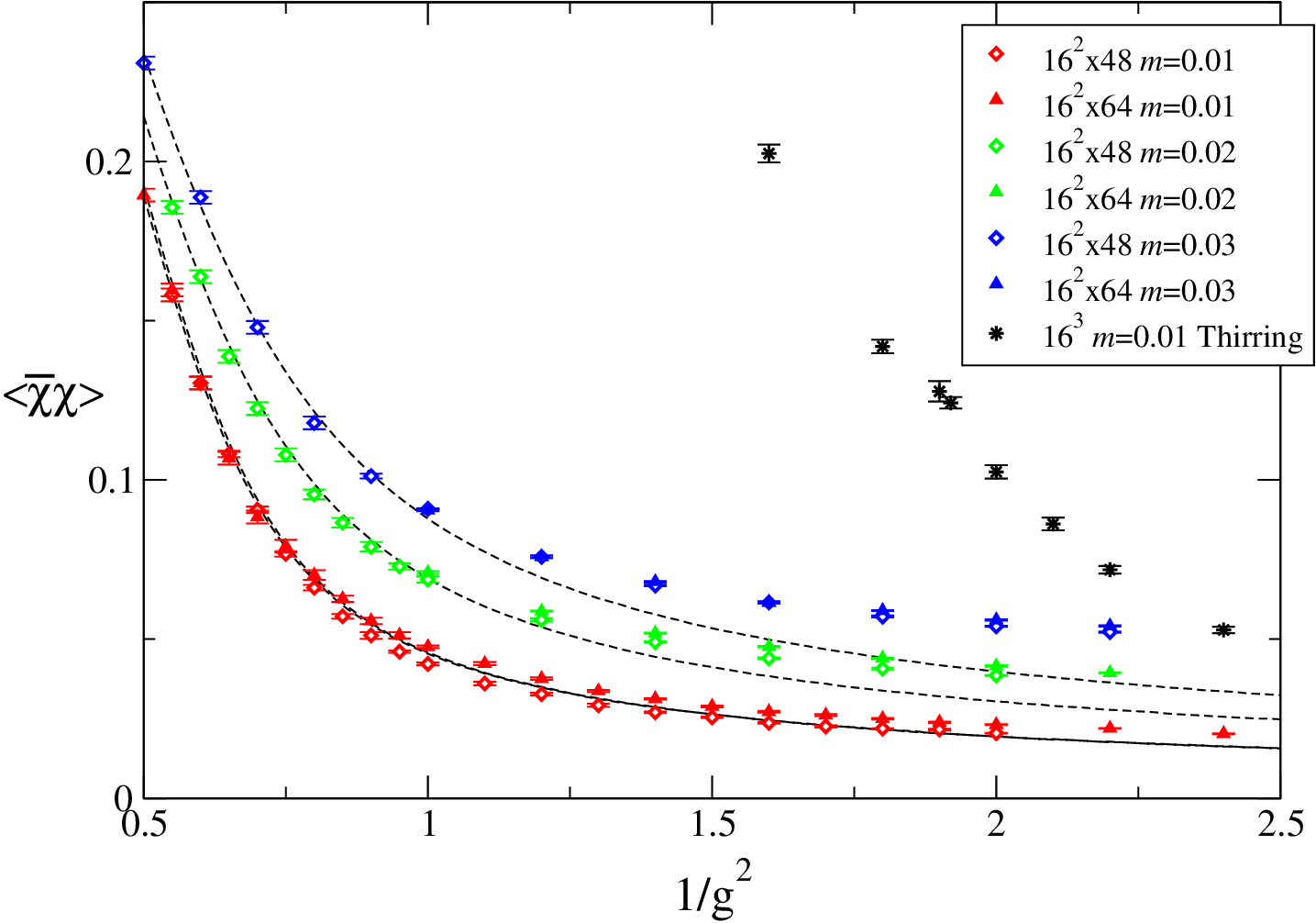}
    \caption{$\langle\bar\chi\chi\rangle$ vs. 
$1/g^2$ for $N_f=2$.} 
   \label{gr:Nf=2}
\end{figure}

Next we turn our attention to the physical case $N_f=2$. Since $N_{fc}>2$, we
expect chiral symmetry to be broken at strong coupling and potentially
restored at some finite $g^2_R$. Accordingly we study
$\langle\bar\chi\chi\rangle$ as a function of $1/g^2$ and here study additional
values of the mass parameter $m$. Our results are summarised in
Fig~\ref{gr:Nf=2}. 

As before, we have attempted to fit a critical equation of state using the 
forms (\ref{eq:power}) to data from $16^2\times64$, and (\ref{eq:fvs}),
in each case replacing $(N_f-N_{fc})$ with $(g^{-2}-g_c^{-2})$. No
stable fits were found unless data with $1/g^2\lapprox1$ were excluded. 
Our most comprehensive fit, using the FVS form and restricting the data to
$L_t\geq48$, $1/g^2\leq0.9$, yields $1/g_c^2=0.632(6)$ and
is shown in the figure. All the fits we found identify a
$1/g^2_c\simeq0.6\gg1/g^2_{\rm peak}$, but all
clearly undershoot the
data at weaker couplings by a considerable margin. 
We conclude that the eqution of state ansatze (\ref{eq:power},\ref{eq:fvs})
are inadequate to describe the data.

\begin{figure}[!hbtp]
    \centering
    \includegraphics[width=13.0cm]{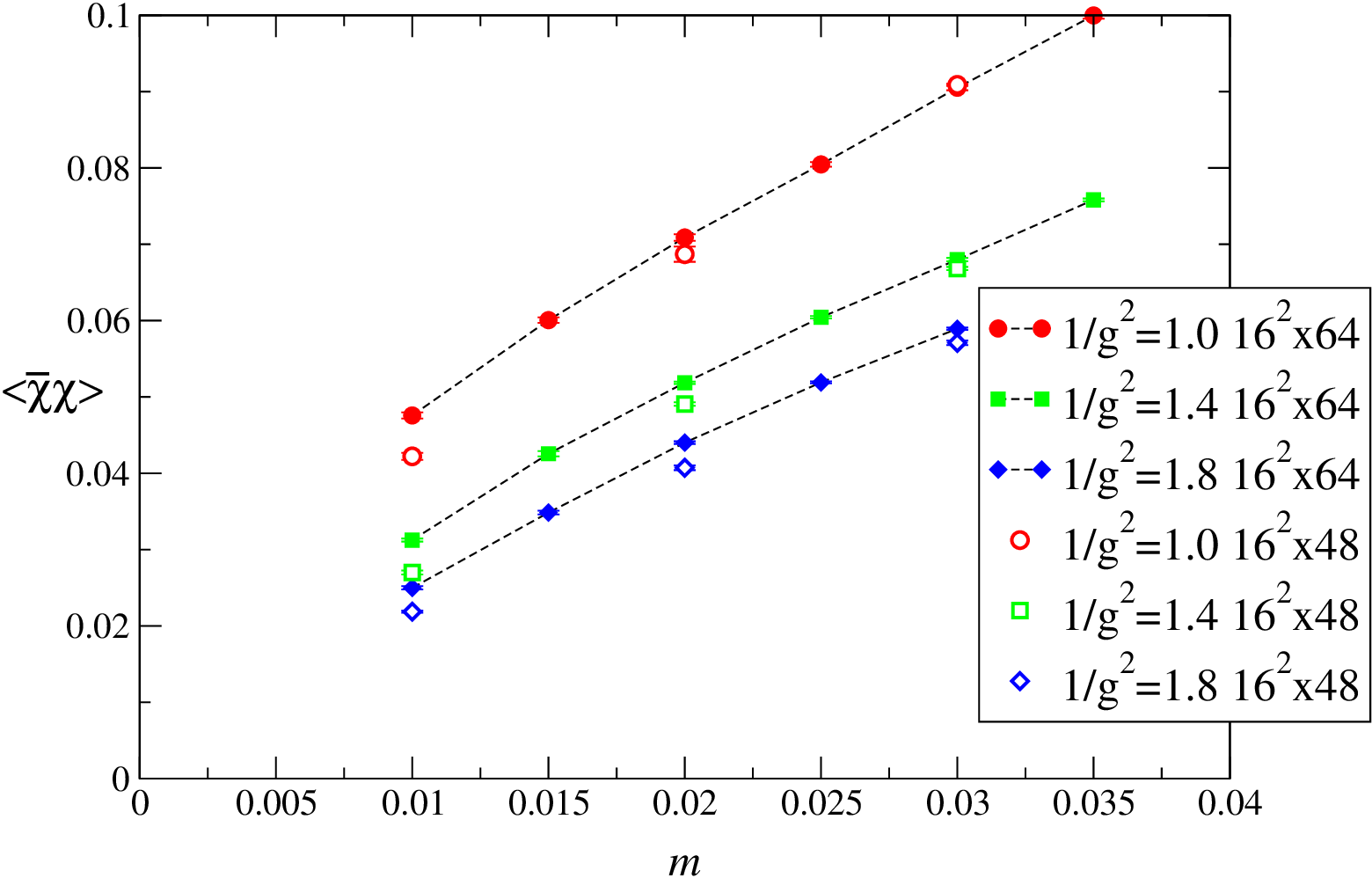}
    \caption{$\langle\bar\chi\chi\rangle$ vs. 
$m$ for $N_f=2$ in the weak-coupling regime.} 
   \label{gr:mass_scaling}
\end{figure}

Instead, we will distinguish between a ``strong coupling'' regime
$1/g^2\lapprox0.75$ where 
$\langle\bar\chi\chi\rangle$ is numerically large and finite volume effects 
are negligible and a ``weak coupling'' regime $1/g^2\gapprox0.75$ where the
opposite holds true. 
Two comments about the weak coupling regime are worth making. First, as is clear
from
Figs.~\ref{gr:Nf=2} and \ref{gr:mass_scaling}, 
finite volume effects are unexpectedly large, and indeed
increase in relative importance until $1/g^2\gapprox1$; it is this
feature which has made a global FVS fit impossible. In a conventional
symmetry-breaking scenario by contrast 
one expects the finite volume effects to be larger in
the broken phase, where there are long-range correlations due to Goldstone
bosons. 
Secondly, in  a chirally symmetric phase one expects
$\langle\bar\chi\chi\rangle\propto m$ for small $m$ and weak interactions;
inspection of the data 
plotted in Fig.~\ref{gr:mass_scaling} appears to imply either that a linear
extrapolation to $m\to0$ would yield a non-vanishing order parameter, or
alternatively, that chiral symmetry restoration for these values of $g^2<g_c^2$
would require $\langle\bar\chi\chi(m)\rangle$ to exhibit some negative
curvature, for which there is some tentative evidence in the figure. 
We conclude that either chiral symmetry remains broken at weak coupling,
or that there are long-range correlations in this region.

We briefly consider alternative scaling scenarios.
In the chiral and thermodynamic limits
two other kinds of behaviour are possible to envisage, and 
are not currently excluded by our data. Firstly, a form
$\langle\bar\chi\chi\rangle=Ae^{-B/g^2}$  would predict broken
chiral symmetry for all $g^2$. This is
barely credible, since the $\chi$ - $\bar\chi$ forces in this model are
weaker than for the $N_f=2$ Thirring model, 
where two independent simulation studies 
find a second
order chiral restoring transition at 
$1/g^2\simeq1.9$~\cite{DelDebbio:1997dv,Barbour:1998yc}
(Fig.~\ref{gr:Nf=2} also shows
Thirring data from $16^3$~\cite{DelDebbio:1997dv}).
Second,
$\langle\bar\chi\chi\rangle=Ae^{-B/(g^{-2}_c-g^{-2})^q}$ describes
chiral symmetry restoration via an infinite order phase transition 
at $g^2=g_c^2$. Without a reliable finite volume scaling hypothesis
we cannot estimate $g^2_c$,
$B$ or $q$. By analogy with the Kosterlitz-Thouless
transition in 2$d$ systems, though,
this scenario predicts a critical and hence
strongly fluctuating system for $g^2>g_c^2$ which could plausibly account for
the observations reported above.

\section{Discussion}

In this paper we studied a model (\ref{eq:Thirgr}) which has very similar
properties, including the same global symmetries, as the graphene-related model
rececently proposed to describe quantum critical behaviour in the $(N_f,e^2)$
plane~\cite{Son:2007ja}.  Using a simulation strategy devised for the 2+1$d$
Thirring model~\cite{Christofi:2007ye}, we have identified the critical number
of flavors separating insulating from conducting phases in the strong coupling
limit  to be $N_{fc}=4.8(2)$. This implies that the strong coupling limit of
graphene with $N_f=2$ is an insulator.  Since the properties of a critical point
should be universal, we expect this result to be a robust prediction of our
work, thus furnishing the first systematic non-perturbative prediction of
quantum critical behaviour in this system.  We also managed a reasonable fit of
our strong-coupling data to an equation of state describing a continuous phase
transition at the critical point, and obtained estimates for the critical
exponents.

The fitted value of $\nu_t$ gives partial information about the nature of
correlations in the vicinity of the fixed point. Substituting our favoured
values $\delta=5.5$, $p=0.86$, $\nu_t=2.1$ into (\ref{eq:BW1.10}), we obtain
$\nu_s={1\over2}\gamma-0.83$. Without studying the order parameter
susceptibility we have no independent estimate of $\gamma$, but note that 
it would need to have a value of $O(6)$ in order for $\nu_s$ to exceed $\nu_t$.
Now, there is a further relation governing the scaling of critical correlation
functions \cite{BW}:
\begin{equation}
(d-2+\eta_t)\nu_t=(d-2+\eta_s)\nu_s\;\;\;\Rightarrow\;\;\;
{\nu_t\over\nu_s}={{1+\eta_s}\over{1+\eta_t}},
\end{equation}
where order parameter correlations
$\langle\bar\chi\chi(0)\bar\chi\chi(x_{s,t})\rangle\propto 
x_{s,t}^{-\eta_{s,t}}$ at 
criticality, with the exponent taking the appropriate value depending on whether
the displacement $x$ is timelike or spacelike. The ratio $\eta_s/\eta_t>1$ if
$\nu_s/\nu_t<1$ and {\it vice versa\/}. However, $\eta_s/\eta_t$ 
may be identified with
the dynamical critical exponent $z$ characterising the quantum critical point,
in the sense that the dynamics remains invariant under the scale
transformation $\vec x\to\ell\vec x;\;x_0\to\ell^zx_0$. By 
considering the anomalous dimension of the Fermi velocity using the $1/N_f$
expansion, Son~\cite{Son:2007ja}
has obtained $z<1$, which has implications for the stability of the
quasiparticle excitations. If we assume that the same critical exponent 
governs both quasiparticle and order parameter correlations, then reconciling 
the two calculations requires an unusually large value of $\gamma$.

Next, we set $N_f$ to the physical value 2 and studied the chiral order
parameter
as a function of coupling strength. Here our results are harder to
interpret; we observe a crossover from strong- to weak-coupling behaviour at
$1/g^2\simeq0.75$, but were unable to model the equation of state,
leaving the nature of the weak-coupling regime unclear. There is evidence, both
from the large finite volume effects and the curvature in $\langle
\bar\chi\chi(m)\rangle$, for strong correlations.
Work is currently in progress 
to study the quasiparticle propagator in order to explore
the dispersion relation and expose any quantum critical behaviour from an
independent direction, and also to further investigate the nature of the
fluctuations in the weak-coupling phase at $N_f=2$.

\section{Acknowledgements}
SJH thanks the Galileo Galilei Institute for Theoretical Physics for hospitality
and the INFN for partial support during the completion of this work.

\end{document}